\documentclass[prd,11pt, amsmath, amsymb, aps, reprint, tightenlines, nofootinbib, longbibliography, abbrv, preprintnumbers]{revtex4-1}

\usepackage[dvipdfmx]{graphicx}
\usepackage[colorlinks=true,linkcolor=blue,citecolor=blue, urlcolor=blue]{hyperref} 
\usepackage{bm,braket}
\usepackage{fancybox, color, float}

\begin{document}
\title{
Quantum dissipation of a heavy quark from a nonlinear stochastic Schr\"odinger equation
}

\author{Yukinao Akamatsu}
\email{akamatsu@kern.phys.sci.osaka-u.ac.jp}
\affiliation{Department of Physics, Osaka University, Toyonaka, Osaka 560-0043, Japan}

\author{Masayuki Asakawa}
\email{yuki@phys.sci.osaka-u.ac.jp}
\affiliation{Department of Physics, Osaka University, Toyonaka, Osaka 560-0043, Japan}

\author{Shiori Kajimoto}
\email{kajimoto@kern.phys.sci.osaka-u.ac.jp}
\affiliation{Department of Physics, Osaka University, Toyonaka, Osaka 560-0043, Japan}

\author{Alexander Rothkopf}
\email{rothkopf@thphys.uni-heidelberg.de}
\affiliation{Institute for Theoretical Physics, Heidelberg University, 69120 Heidelberg, Germany}
\affiliation{Faculty of Science and Technology, University of Stavanger, 4036 Stavanger, Norway}

\begin{abstract}
We study the open system dynamics of a heavy quark in the quark-gluon plasma using a Lindblad master equation.
Applying the quantum state diffusion approach by Gisin and Percival, we derive and numerically solve a nonlinear stochastic Schr\"odinger equation for wave functions, which is equivalent to the Lindblad master equation for the density matrix.
From our numerical analysis in one spatial dimension, it is shown that the density matrix relaxes to the Boltzmann distribution in various setups (with and without external potentials), independently of the initial conditions.
We also confirm that quantum dissipation plays an essential role not only in the long-time behavior of the heavy quark but also at early times if the heavy quark initial state is localized and quantum decoherence is ineffective.
\end{abstract}

\date{\today}

\maketitle

\section{Introduction}
The study of phases of strongly interacting matter under extreme conditions \cite{Akiba:2015jwa, Fukushima:2010bq} is attracting broad interest in recent years, extending far beyond nuclear matter.
Fruitful interdisciplinary collaborations, among others, with the field of ultracold atomic gases (for a review see e.g. \cite{Adams:2012th}) have broadened the scope of how to explore the origins of the universe. 

One example of extreme conditions is very high temperatures, which are particularly interesting, since they bear direct relevance to the birth and early history of our universe.
The properties of the hottest matter ever created on the Earth are being investigated at current collider facilities, such as the Relativistic Heavy-Ion Collider (RHIC) and the Large Hadron Collider (LHC) and upcoming facilities, such as NICA and FAIR. There, two heavy nuclei are collided at ultra-relativistic energies in order to create a novel deconfined state of matter known as the quark-gluon plasma (QGP) \cite{Jacak:2012dx}.

The transition from hadronic matter to the QGP is characterized by the liberation of colored degrees of freedom (quarks and gluons) otherwise confined inside hadrons.
This qualitative picture is confirmed by lattice QCD calculations of, for example, the QCD entropy, which show a significant rise around the pseudocritical temperature \cite{Borsanyi:2013bia, Bazavov:2014pvz,Borsanyi:2016ksw,Bazavov:2017dsy}. 

In analogy with the Debye screening phenomenon in electromagnetic plasmas, the liberated colored particles may rearrange themselves around a test color charge such that the test charge is screened with a finite screening length \cite{McLerran:1981pb}.
Compared to the confining string-like force between color sources in the vacuum, the in-medium force in a high temperature QGP is hence drastically modified and becomes short ranged \cite{Kaczmarek:2005ui,Maezawa:2007fc,Bazavov:2016uvm}.

The in-medium modification of the QCD force has been expected to have dramatic consequences on the behavior of bound states of heavy quarks and antiquarks, so called heavy quarkonium \cite{Brambilla:2010cs,Andronic:2015wma}.
In turn the dynamics of quarkonium states observed in heavy-ion collisions promises a direct window on the phase structure of the bulk matter, in which they are immersed.

One classic prediction in this context is the enhanced probability of dissociation of charm quark pairs in nuclear collisions, in case that a QGP is created \cite{Matsui:1986dk}.
The qualitative behavior of $J/\psi$ and $\Upsilon$ yields observed at RHIC and $\Upsilon$ yields at LHC support this idea \cite{Adamczyk:2013poh, Adare:2014hje, Adare:2006ns, Adare:2008sh, Adare:2011yf, Adamczyk:2013tvk, Chatrchyan:2011pe, Chatrchyan:2012lxa, Khachatryan:2016xxp, Abelev:2014nua}.

However, the enhanced production rate of charm quark pairs at LHC complicates an interpretation of $J/\psi$ yields at LHC \cite{Abelev:2013ila, Adam:2016rdg}.
The reason is that now another process to create $J/\psi$ needs to be taken into account, i.e. the  non-negligible probability that initially uncorrelated charm quarks end up forming bound states at the phase boundary \cite{BraunMunzinger:2000px}, i.e. in the late stages of the collision at freezeout.
The successful prediction of $J/\psi$ yields at the LHC by means of the statistical model of hadronization \cite{Andronic:2010dt} is seen as support for the existence of such a production mechanism.
 
At present it is still not clear at which collision energy the two effects, i.e. the suppressed and enhanced yields of quarkonium become comparable.
In order to more clearly interpret the collected experimental results we thus need to better understand the dynamics of quarkonia in the QGP.
I.e.\ the development of a unified quantum mechanical description for the real-time equilibration of quarkonia is called for.

Recently, the dynamics of heavy quark pairs in the QGP has been actively studied by various methods \cite{Aarts:2016hap}.
Among them are kinetic descriptions \cite{Rapp:2008tf, Zhao:2010nk, Zhao:2011cv, Zhou:2014kka, Song:2011nu, Emerick:2011xu, Zhou:2014hwa, Yao:2017fuc} or dynamical models involving a complex potential \cite{Strickland:2011mw, Strickland:2011aa, Krouppa:2015yoa, Krouppa:2016jcl, Krouppa:2017jlg} for quarkonium.
One more recent promising approach is the open quantum system
formulation for the heavy quark pairs \cite{Young:2010jq,
Borghini:2011ms, Akamatsu:2011se, Rothkopf:2013kya, Akamatsu:2014qsa,
Akamatsu:2015kaa, Kajimoto:2017rel, Blaizot:2015hya, Blaizot:2017ypk,
Blaizot:2018oev, Brambilla:2016wgg, Brambilla:2017zei, DeBoni:2017ocl,
Katz:2015qja}, which developed concurrently to computations of the in-medium complex potential \cite{Laine:2006ns, Beraudo:2007ky, Brambilla:2008cx, Rothkopf:2011db, Burnier:2014ssa, Burnier:2015tda,  Burnier:2016mxc}.

In the open quantum system formulation \cite{BRE02}, we distinguish the
subsystem of interest (quarkonium) from the environment (QGP), which is made possible by a hierarchy of time scales in each sector.
The dynamics of the environment is fast enough, so that we can trace it out and replace its coupling to the subsystem by the average response to a slowly changing external source. 

In this way a master equation for the reduced density matrix of a heavy quark pair can be obtained, which can be expressed as arising from the contribution of three kinds of forces: 
the screened potential, thermal fluctuations, and dissipation \cite{Akamatsu:2014qsa}.
The first two forces combine into a fluctuating potential force
(stochastic potential \cite{Akamatsu:2011se, Rothkopf:2013kya,
Kajimoto:2017rel}), while the last one is an irreversible force.
So far there exist only a few numerical analyses of the quantum dissipation of quarkonium in the QGP \cite{DeBoni:2017ocl, Brambilla:2017zei, Katz:2015qja}.
However, quantum dissipation is an essential ingredient to understand the long-time behavior of a heavy quark pair in the QGP.
It is particularly important to know how and when quantum dissipation influences the time evolution of the heavy quark pair because the lifetime of the QGP in heavy-ion collisions is not long enough for the heavy quark pair to get fully equilibrated.

In this paper we consider as a first step the physics of a single heavy quark immersed in a hot medium.
We numerically solve the corresponding master equation and study its equilibrium solution as well as the effects of dissipation. 
The central conceptual result of this study is a stochastic unravelling prescription for the master equation, in which the wave function is evolved in terms of a stochastic Schr\"odinger equation and the mixed states of the density matrix emerge from an ensemble average.

Our results are obtained following the  ``Quantum State Diffusion" approach developed by Gisin and Percival \cite{gisin1992quantum, percival1998quantum} and by subsequently solving the corresponding nonlinear stochastic Schr\"odinger equation.
Note that by applying the Quantum State Diffusion approach to a heavy quark master equation derived in the Lindblad form \cite{Akamatsu:2014qsa, lindblad1976generators} (see Section \ref{sec:QSD} for the definition) we obtain for the first time in this context a non-linear Schr\"odinger equation with a clear connection to the underlying microscopic theory. 

For simplicity, we consider a heavy quark in the QGP in one spatial dimension.
It is then shown that the master equation possesses a steady state solution consistent with the Boltzmann distribution $\rho_{\rm eq}\propto e^{-\beta H}$.
Furthermore, we observe that the approach to equilibrium depends on initial conditions and cannot be captured by a single decay rate.
Finally, we analyze the effect of quantum dissipation by comparing with simulations in which the dissipative terms are dropped. 
Their effect, as expected, is essential at later times but interestingly already plays an important role at rather early times if the initial wave function is well localized and decoherence by thermal fluctuations is ineffective.

This paper is organized as follows. In Sec.~\ref{sec:QSD}, we introduce the Quantum State Diffusion approach applicable to a general Lindblad master equation.
We then apply this approach to the Lindblad equation for a single heavy quark in the quark-gluon plasma at high temperature and derive the corresponding nonlinear stochastic Schr\"odinger equation.
In Sec.~\ref{sec:results}, we solve this nonlinear stochastic Schr\"odinger equation numerically and analyze the relaxation process of the heavy quark.
In addition we study the effect of quantum dissipation on the time evolution of a heavy quark, before we summarize our work in Sec.~\ref{sec:conclusion}.

\section{Quantum state diffusion for open quantum systems}
\label{sec:QSD}
\subsection{Lindblad equation and quantum state diffusion}
There is a particularly useful class of master equations for open quantum systems, which is Markovian and fulfills basic physical requirements: the reduced density matrix $\rho$ is hermitian ($\rho = \rho^{\dagger}$), correctly normalized (${\rm Tr}\rho = 1$), and positive ($\langle\alpha|\rho|\alpha\rangle\geq 0$ for any state $|\alpha\rangle$) during its time evolution.
Such master equations may be written in general as
\begin{align}
\label{eq:Lindblad}
\frac{d}{dt}\rho(t) = -i\left[H,\rho\right]
+\sum_n\left(
2L_n\rho L_n^{\dagger} - L_n^{\dagger}L_n\rho - \rho L_n^{\dagger} L_n
\right),
\end{align}
in the so called Lindblad form \cite{lindblad1976generators}.
Here the evolution of the density matrix operator is described in terms of the full dimension of the system Hilbert space.
This makes a direct numerical simulation computationally highly demanding, especially in realistic 3+1 dimensions.
There are, however, several ways to solve the Lindblad equation by what is known as stochastic unravelling.
I.e. by carrying out a stochastic evolution of the wave functions of the system instead, whose ensemble average then correctly reproduces the density matrix.
One such stochastic unravelling corresponds to the quantum state diffusion (QSD) approach \cite{gisin1992quantum}.

Given the Lindblad equation \eqref{eq:Lindblad}, the corresponding QSD equation is a stochastic nonlinear Schr\"odinger equation:
\begin{align}
\label{eq:qsd}
&|d\psi\rangle = |\psi(t+dt)\rangle - |\psi(t)\rangle \nonumber \\
&=-i H|\psi(t)\rangle dt
+\sum_n\left(
\begin{aligned}
&2\langle L_n^{\dagger}\rangle_{\psi} L_n - L_n^{\dagger}L_n  \\
&- \langle L_n^{\dagger}\rangle_{\psi}\langle L_n\rangle_{\psi}
\end{aligned}
\right) |\psi(t)\rangle dt \nonumber \\
& \quad +\sum_n \left(L_n - \langle L_n\rangle_{\psi}\right)|\psi(t)\rangle d\xi_n,
\end{align}
with complex white noises $d\xi_n$ whose mean and variance are given by
\begin{subequations}
\begin{align}
{\rm M}\left( d\xi_n\right)&={\rm M}\left( \Re (d\xi_n)\Im (d\xi_m)\right) = 0, \\
{\rm M}\left(\Re (d\xi_n)\Re (d\xi_m)\right)
&={\rm M}\left( \Im (d\xi_n)\Im (d\xi_m)\right) = \delta_{nm} dt.
\end{align}
\end{subequations}
Here $\langle O \rangle_{\psi}\equiv \langle\psi| O |\psi\rangle$ denotes the quantum expectation value of an operator with respect to a state $\psi$ and ${\rm M}(O)$ denotes the statistical average of $O$.

This stochastic evolution equation is solved in the It$\hat{\rm o}$ discretization scheme.
By using $\psi(t)$ everywhere in Eq.~\eqref{eq:qsd} we obtain the wave function at the next discrete time step $\psi(t+dt)$.
In the limit $dt\to 0$, the QSD equation \eqref{eq:qsd} preserves the norm of $\psi$ in each stochastic update.
The initial wave function is distributed according to the initial mixed (or pure) state of the density matrix.
The density matrix is subsequently constructed from an ensemble average of wave functions $\psi(t)$,
\begin{align}
\rho(t) = {\rm M}\left(|\psi(t)\rangle\langle\psi(t)|\right),
\end{align}
and obeys the Lindblad equation \eqref{eq:Lindblad} in the $dt\to 0$ limit.

The QSD equation can also be formulated for unnormalized wave functions $\phi$
\begin{align}
\label{eq:qsd2}
& |d\phi\rangle = |\phi(t+dt)\rangle - |\phi(t)\rangle \nonumber\\
&= -i H|\phi(t)\rangle dt
+\sum_n\left(
2\langle L_n^{\dagger}\rangle_{\phi} L_n
- L_n^{\dagger}L_n\right) |\phi(t)\rangle dt \nonumber\\
& \quad +\sum_n L_n |\phi(t)\rangle d\xi_n,
\end{align}
with the same complex noise.
Here $\langle O \rangle_{\phi} \equiv \langle\phi| O |\phi\rangle/\langle\phi|\phi\rangle$ denotes the quantum expectation value. The density matrix constructed by
\begin{align}
\rho(t)= {\rm M}\left(\frac{|\phi(t)\rangle\langle\phi(t)|}{\langle\phi(t)|\phi(t)\rangle}\right)
\end{align}
is again a solution of the master equation \eqref{eq:Lindblad}.
In our numerical simulation, we implement eq.\eqref{eq:qsd2}.

\subsection{Quantum state diffusion for a heavy quark in the quark-gluon plasma}
Let us now consider the theory of open quantum systems for a single heavy quark in the QGP and stochastically unravel its Lindblad equation via the QSD approach.
The Lindblad equation for heavy quarks has been derived in
\cite{Akamatsu:2014qsa} by treating the scattering between heavy quarks
and medium particles perturbatively, i.e. by assuming that the QCD coupling constant $g$ is small.
It is further assumed that the heavy quark mass $M$ is much larger than the temperature $T/M\ll 1$ so that there exists a time scale hierarchy between heavy quarks and medium particles. 

The Lindblad master equation for a single heavy quark is given by the following operators \cite{Akamatsu:2014qsa}
\begin{subequations}
\label{eq:QCDLindblad}
\begin{align}
H &= -\frac{\nabla^2}{2M} + V_{\rm ext}(x), \\
L_{k} &= \sqrt{\frac{\tilde D(k)}{2V}} e^{i\bm k \cdot \bm x/2}\left(1+\frac{i\bm k \cdot\bm\nabla}{4MT}\right)e^{i\bm k \cdot \bm x/2},\\
\tilde D(k) &= g^2 T\frac{\pi m_D^2}{k(k^2 + m_D^2)^2}, \quad
m_D = gT\sqrt{\frac{N_c}{3} + \frac{N_f}{6}},
\end{align}
\end{subequations}
with $N_c$ and $N_f$ being the numbers of colors and quark flavors, respectively.
The Lindblad operator $L_k$ describes the scattering process between a heavy quark and medium particles with momentum transfer $\bm k$, taking place with rate $\tilde D(k)$.
The term $\propto e^{i\bm k \cdot\bm x}$ in $L_k$ describes thermal fluctuations, while the term $\propto e^{i\bm k \cdot\bm x/2}\frac{i\bm k\cdot\bm\nabla}{4MT}e^{i\bm k \cdot\bm x/2}$ describes dissipation and originates in the recoil of the heavy quark during the collision. For simplicity, we ignore the effects of internal color degrees of freedom
\footnote{
Also, we assume that it is admissible to set the second order coefficient in the derivative expansion of the Feynman-Vernon influence functional $\tilde A(k) = \tilde D(k)/8T^2$. This reduces the number of Lindblad operators that need to be considered.
}.

We can calculate the parts of the QSD equation as follows.
The nonlinear term is given by
\begin{align}
\label{eq:qsdhq1}
&2\sum_{k}\langle L_{k}^{\dagger}\rangle_{\phi}L_{k}\phi(x)
= \frac{1}{\int d^3 y |\phi(y)|^2} \\
&\times \int d^3 y
\left[n_{\phi}(y) f(x-y) +\frac{i}{4T}\bm j_{\phi}(y) \cdot \bm g(x-y) \right]\phi(x), \nonumber
\end{align}
where $n_{\phi}$ and $\bm j_{\phi}$ denote the probability density and current:
\begin{subequations}
\label{eq:nandj}
\begin{align}
n_{\phi}(x) &\equiv \phi^*(x)\phi(x), \\
\bm j_{\phi}(x) &\equiv \frac{1}{2iM} \left[\phi^*(x)\bm\nabla\phi(x) - \left(\bm\nabla\phi^*(x)\right) \phi(x)\right],
\end{align}
\end{subequations}
and $f$ and $g_i$ are operators defined as
\begin{subequations}
\label{eq:fandg}
\begin{align}
f(x-y) &\equiv \left(1+ \frac{\nabla^2}{8MT}\right)D(x-y) + \frac{\bm \nabla D(x-y)}{4MT}\cdot\bm\nabla_x,\\
g_i(x-y) &\equiv \left(1+ \frac{\nabla^2}{8MT}\right)\nabla_i D(x-y) \nonumber \\
 & \quad + \frac{\bm \nabla \nabla_i D(x-y)}{4MT}\cdot \bm\nabla_x. 
\end{align}
\end{subequations}
The function $D(x)$ in the equations above is the inverse Fourier transform of $\tilde D(k)$.
The linear deterministic term reads
\begin{align}
\label{eq:qsdhq2}
\sum_k L_k^{\dagger} L_k\phi(x)
=&\frac{1}{2}
\left(D(0)+\frac{\nabla^2 D(0)}{4MT} + \frac{\nabla^4 D(0)}{64M^2T^2}\right) \phi(x) \nonumber \\
&+ \frac{\nabla_i\nabla_j D(0)}{32M^2T^2}\nabla_i\nabla_j\phi(x),
\end{align}
and the linear stochastic term amounts to
\begin{align}
\label{eq:qsdhq3}
&\sum_k L_k\phi(x) d\xi_k \nonumber \\
& \quad =\left[d\zeta(x)+\frac{\nabla^2d\zeta(x)}{8MT}+ \bm\nabla d\zeta(x)\cdot \frac{\bm \nabla}{4MT}
\right]\phi(x),
\end{align}
where the definition and the correlation of the complex noise field $d\zeta(x)$ is given by
\begin{subequations}
\begin{align}
&d\zeta(x)\equiv \sqrt{\frac{V}{2}} \int \frac{d^3 k}{(2\pi)^3} \sqrt{\tilde D(k)} e^{i\bm k\cdot \bm x}d\xi_k, \\
&{\rm M}\left(d\zeta(x)d\zeta^*(y)\right) = D(x-y)dt, \\
&{\rm M}\left(d\zeta(x)d\zeta(y)\right) = {\rm M}\left(d\zeta^*(x)d\zeta^*(y)\right) = 0.
\end{align}
\end{subequations}
Using Eqs.~\eqref{eq:qsdhq1}, \eqref{eq:qsdhq2}, and \eqref{eq:qsdhq3}, we may now perform the QSD simulation for a single heavy quark.

\section{Results of numerical simulation}
\label{sec:results}
In this section we explicitly check the application of the QSD approach and investigate the properties of the Lindblad equation in three simple settings.
We consider a single heavy quark in the QGP in one spatial dimension either with or without external potentials.

In particular, we study the equilibration of the heavy quark in each setting and discuss the importance of quantum dissipation.
As external potentials we deploy either the harmonic potential or the regularized Coulomb potential:
\begin{align}
\label{eq:potential}
V_{\rm ext}(x) = \frac{1}{2}M\omega^2x^2, \quad -\frac{\alpha}{\sqrt{x^2 + r_c^2}}.
\end{align}
where $r_c=1/M$. The noise correlation function $D(x)$ is set to have correlation length $\sim 1/m_D$ and is approximated with a Gaussian dependence on distance
\begin{align}
\label{eq:dgaussian}
D(x) = \gamma \exp\left[-x^2/l_{\rm corr}^2\right].
\end{align}
The parameters of our numerical setup are summarized in Table \ref{tbl:setup}.

The Hamiltonian time evolution is solved by the fourth-order Runge-Kutta method and that for the other parts is implemented via an explicit forward step according to the QSD equation \eqref{eq:qsd2} with $dt=\Delta t$.
We check that the discretization effects for both $\Delta t$ and $\Delta x$ are negligible by comparing with results with smaller $\Delta t$ or $\Delta x$.
In the simulation periodic boundary conditions are employed so that the noise correlation is replaced with
\begin{subequations}
\begin{align}
&{\rm M} \left(d\zeta(x)d\zeta^*(y)\right) = D(r_{xy})\Delta t , \\
& r_{xy} = \min \{ |x-y|, N_x\Delta x - |x-y|\},
\end{align}
\end{subequations}
and the function $D(x-y)$ in the QSD equation \eqref{eq:fandg} is also replaced by $D(r_{xy})$.
We confirm by changing $N_x$ that the volume $N_x\Delta x/l_{\rm corr} \simeq 13$ is large enough so that we can neglect finite volume effects.

\begin{table}[b]
\caption{Numerical setup and parameters of the potentials.
$N_x=128 (127)$ is used for the harmonic (regularized Coulomb) potential.}
\label{tbl:setup}
\begin{center}
\begin{tabular}{ccc}
\hline
\hline
$\Delta x$ & $\Delta t$ & $N_x$  \\ \hline
\ $1/M$ \ & \ $0.1M(\Delta x)^2$ \ & \ 128, 127 \ \\ \hline
\end{tabular}
\end{center}
\begin{center}
\begin{tabular}{ccc|ccc}
\hline
\hline
$T$ & $\gamma$ & $l_{\rm corr}$ & $\omega$ & $\alpha$ & $r_c$ \\ \hline
 \ $0.1M$ \ & \ $T/\pi$ \ & \ $1/T$ \ & \ $0.04M$ \ & \ 0.3 \ & \ $1/M$ \ \\ \hline
\end{tabular}
\end{center}
\end{table}

\subsection{Equilibration of a heavy quark}
The Lindblad equation itself does not guarantee the Boltzmann distribution $\rho\propto \exp(-H/T)$ to be the static solution (see Appendix \ref{app:steadystate}).
In the derivation of the Lindblad equation, the fluctuation-dissipation theorem for the environment, i.e. the QGP sector, constrains the terms implementing the fluctuation and dissipation of the heavy quarks.
To be specific, the coefficient $i/4MT$ in $L_k$ is determined from the fluctuation-dissipation theorem for a thermal QGP medium.
Therefore, we expect that the equilibrium density matrix is close to the Boltzmann distribution $\rho\propto \exp(-H/T)$.
Here we analyze how well the equilibration of the density matrix is achieved and study its equilibrium properties.

\subsubsection{In the absence of an external potential}
\begin{figure}
\includegraphics[clip, angle=-90, width=0.45\textwidth]{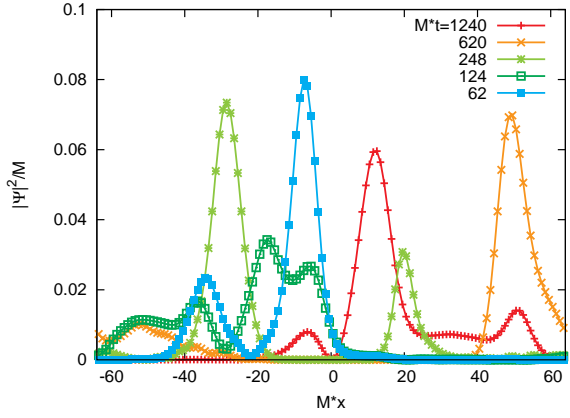}
\caption{
Profiles of a normalized wavefunction at different times in one sample event.
}
\label{fig:wavefunctions}
\end{figure}

\begin{figure}
\includegraphics[clip, angle=-90, width=0.5\textwidth]{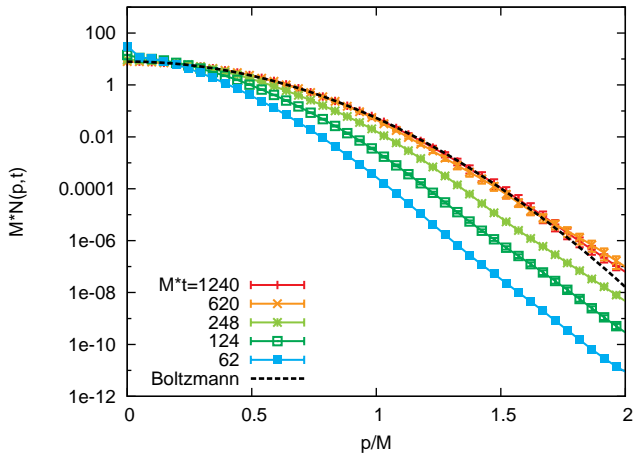}
\caption{
Time evolution of the momentum distribution of a heavy quark.
The bars denote statistical errors.
The dashed line corresponds to the Boltzmann distribution with $T=0.1M$.
}
\label{fig:free_pdist}
\end{figure}

The initial heavy quark wave function is taken to be uniform (plane wave with zero momentum).
In Fig.~\ref{fig:wavefunctions}, we show the profiles of a normalized wavefunction at different times in one sample event.
A wavefunction type typically encountered here is a localized solitonic state, arising from the nonlinearity of the evolution equation.
Figure \ref{fig:free_pdist} on the other hand contains the time evolution of the momentum distribution of the heavy quark:
\begin{subequations}
\begin{align}
N(p,t) &\equiv {\rm M}\left(\frac{|\tilde\phi(p,t)|^2}{\langle\phi(t)|\phi(t)\rangle}\right),\\
\tilde\phi(p,t)&=\int dx e^{-ipx}\phi(x,t),
\end{align}
\end{subequations}
where $p$ takes on values available on a periodic lattice of size $N_x\times \Delta x$.

The corresponding classical dynamics of the heavy quark is a Brownian motion with a drag force
\begin{align}
\frac{dp}{dt} = -\frac{\gamma}{MTl_{\rm corr}^2} p.
\end{align}
Its typical relaxation time is $\tau_{\rm relax} = \frac{MTl_{\rm corr}^2}{\gamma}=100\pi/M$.
We can see that the momentum distribution approaches the Boltzmann distribution with temperature $T=0.1M$ over a time scale $\sim \tau_{\rm relax}$.
Note also that at late times ($t=620/M, 1240/M$) slight deviations from the Boltzmann distribution are observed above $p\agt 1.5M$.
The reason lies in the poor convergence of the gradient expansion, applied in evaluating the Lindblad operators, at high momenta ($p\gtrsim M$), when one takes into account the effects of dissipation.
On the other hand, we should not rely on the nonrelativistic description for a heavy quark with such a high momentum, so this limitation is not so restrictive in practice.

\subsubsection{In the presence of external potentials}
\begin{figure}
\centering
\includegraphics[clip, angle=-90, width=0.65\textwidth]{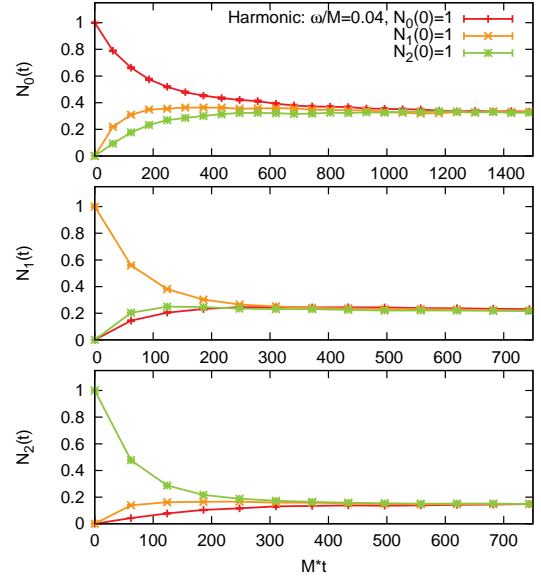}
\caption{
Time evolution of the occupation number of the eigenstates in the harmonic potential with $\omega=0.04M$.
The bars denote statistical errors.
}
\label{fig:harmonic_evolution}
\end{figure}
\begin{figure}
\centering
\includegraphics[clip, angle=-90, width=0.65\textwidth]{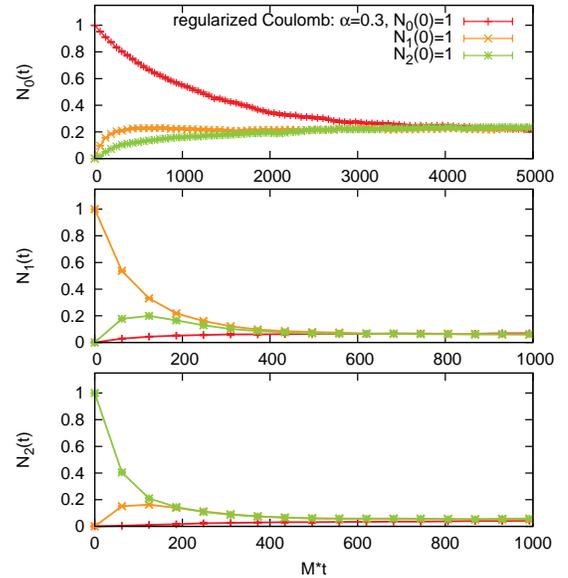}
\caption{
Time evolution of the occupation number of the eigenstates in the regularized Coulomb potential with $\alpha=0.3$ and $r_c=1/M$. 
The bars denote statistical errors.
}
\label{fig:coulomb_evolution}
\end{figure}

Let us turn to the case with external potentials present next.
The potential here is added by hand out of pure theoretical interest and should not be confused with the potential for a quarkonium state.
One should rather think of a single heavy quark in a fictitious trap.

The initial heavy quark wave function is taken to be the ground state, the first, or the second excited states of the corresponding Hamiltonian.
The time evolution of the occupation number of these levels,
\begin{align}
N_i(t)\equiv {\rm M}\left(\frac{|\langle \psi_i|\phi(t)\rangle|^2}{\langle\phi(t)|\phi(t)\rangle}\right), \quad
H|\psi_i\rangle = E_i|\psi_i\rangle,
\end{align}
is shown in Fig.~\ref{fig:harmonic_evolution} for the harmonic potential and in Fig.~\ref{fig:coulomb_evolution} for the regularized Coulomb potential.
Independent of the initial conditions, the occupation numbers converge to their equilibrium values.
In contrast, the relaxation time depends on the initial condition.

One might expect a naive relaxation process
\begin{align}
N_i(t) = (N_i^{\rm ini}-N_i^{\rm eq})\exp(-\Gamma_i t) + N_i^{\rm eq},
\end{align}
to describe the dynamics, as motivated and applied in the rate equation approach to heavy quarks.
However, from Fig.~\ref{fig:harmonic_evolution} and Fig.~\ref{fig:coulomb_evolution} it is obvious that a single decay rate cannot capture the relaxation of the eigenstate occupation.
Note that actually there are cases where the occupation number shows a non-monotonic approach to equilibrium.

\begin{figure}
\centering
\includegraphics[angle=-90, width=0.7\textwidth]{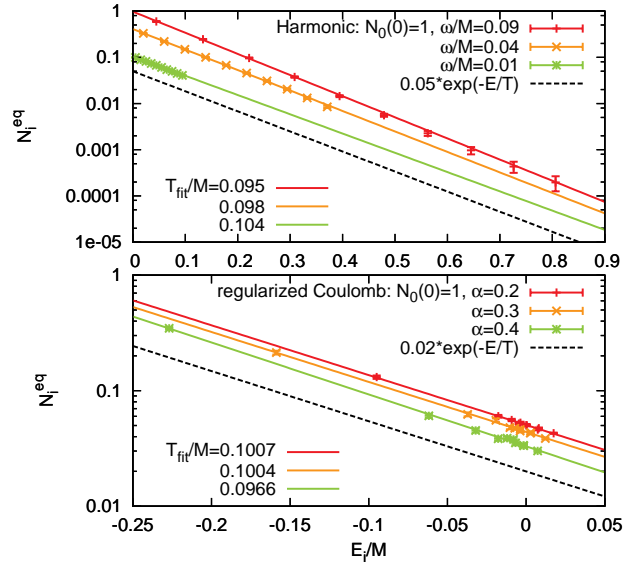}
\caption{
Equilibrium occupation of the lowest ten eigenstates for the harmonic potential (upper panel) and the regularized Coulomb potential (lower panel).
The parameters are varied as $\omega/M=0.01, 0.04, 0.09$ for the harmonic potential and $\alpha=0.2, 0.3, 0.4$ and $r_c=1/M$ for the regularized Coulomb potential.
The bars denote statistical errors.
The data are fitted by $C\cdot\exp(-E_i/T_{\rm fit})$ and the dashed lines indicate the slope of a Boltzmann distribution with $T=0.1M$.
}
\label{fig:potential_equilibrium}
\end{figure}

In order to investigate the properties of the equilibrium density matrix, we show in Fig.~\ref{fig:potential_equilibrium} the equilibrium distribution of the lowest ten levels as a function of the eigenenergy for the harmonic and the regularized Coulomb potentials.
In the figures, we plot the results for several different potential parameters: $\omega/M = 0.01, 0.04, 0.09$ for the harmonic potential and $\alpha = 0.2, 0.3, 0.4$ and $r_c=1/M$ for the regularized Coulomb potential.
The initial condition is chosen to be the ground state of each Hamiltonian.
The equilibrium distribution is calculated at late enough time for each
setup: $Mt=1550, 3100, 4650$ for $\omega/M=0.01, 0.04, 0.09$ and
$Mt=4650, 7750, 9300$ for $\alpha=0.2, 0.3, 0.4$, respectively.
In all cases, the distribution is close to the Boltzmann distribution with $T=0.1M$.
We also calculate the real and imaginary parts of the off-diagonal elements of the density matrix in equilibrium and check that they are consistent with zero within the statistical uncertainty.

\subsection{The effect of dissipation}
\begin{figure}
\centering
\includegraphics[clip, angle=-90, width=0.5\textwidth]{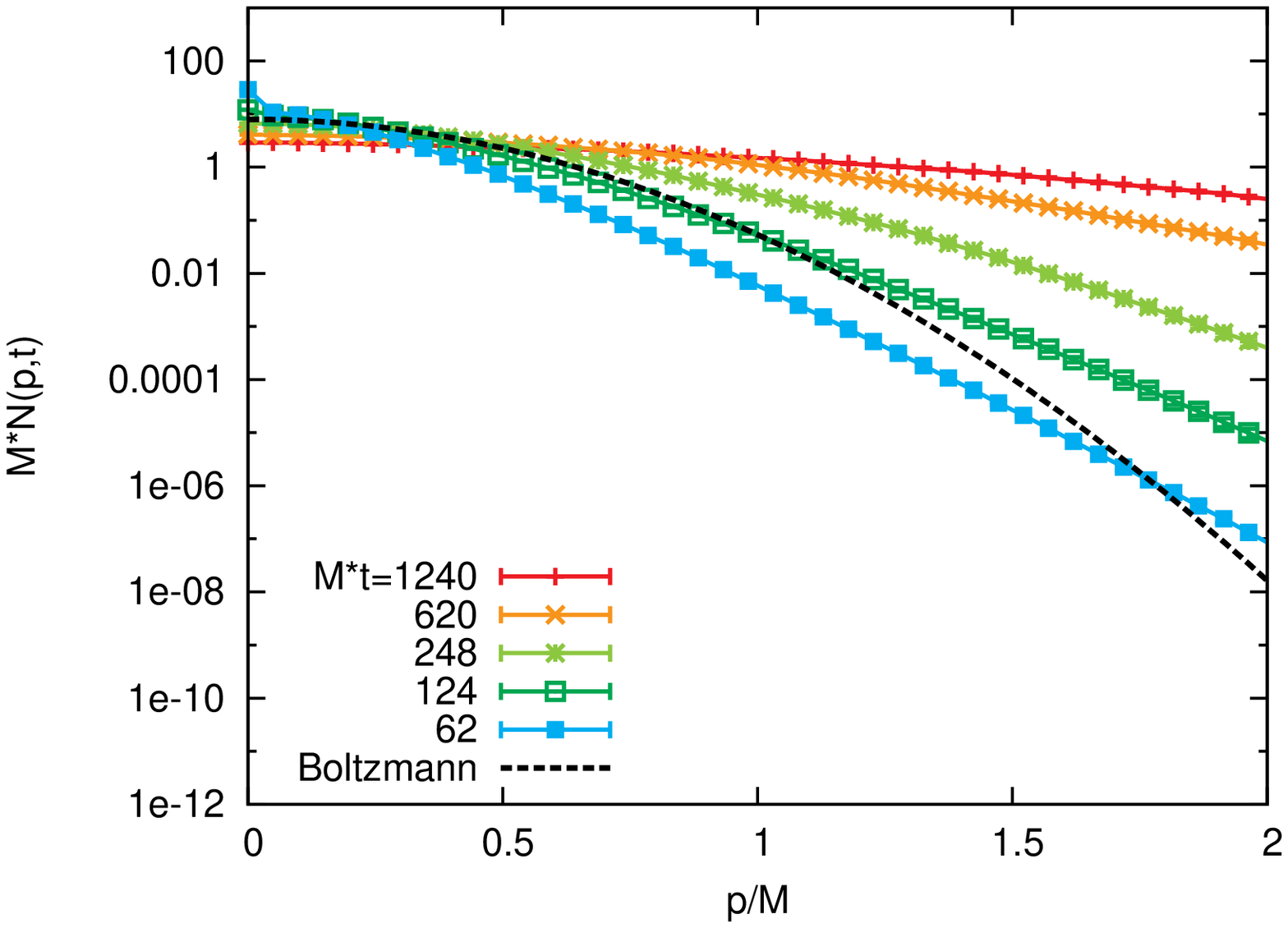}
\caption{
Time evolution of the momentum distribution of a heavy quark without dissipation.
The bars denote statistical errors and the dashed line corresponds to a Boltzmann distribution with $T=0.1M$.
}
\label{fig:free_pdist_nodiss}
\end{figure}

As a last consideration let us now turn off the quantum dissipation.
Since quantum dissipation is described by the term $\propto e^{i\bm k \cdot\bm x/2}\frac{i\bm k\cdot\bm\nabla}{4MT}e^{i\bm k \cdot\bm x/2}$, we can switch it off by taking the $M\to \infty$ limit in the QSD equation everywhere, except in the Hamiltonian part.
In Fig.~\ref{fig:free_pdist_nodiss}, we show the corresponding time evolution of the momentum distribution as given by the QSD equation without dissipation.
Clearly, the distribution does not approach the equilibrium Boltzmann distribution.
Instead the energy gained by the heavy quark from thermal fluctuations is not dissipated back to the system and the heavy quark overheats.

\begin{figure}
\centering
\includegraphics[angle=-90, width=0.65\textwidth]{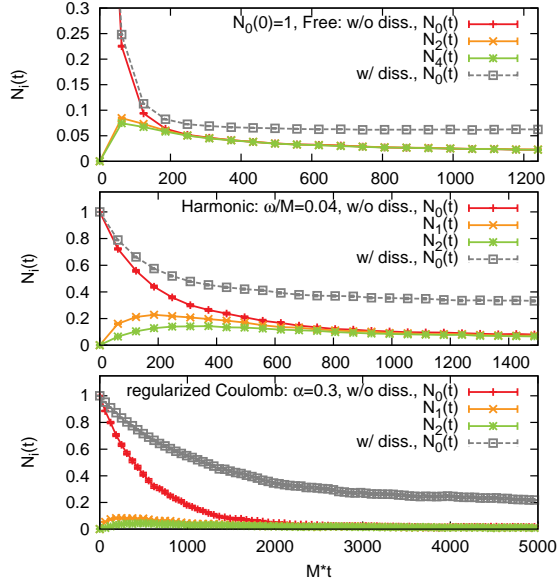}
\caption{
Time evolution of the occupation number of the eigenstates without dissipation.
For comparison, the time evolution with dissipation is also plotted.
The bars denote statistical errors.
}
\label{fig:potential_evolution_nodiss}
\end{figure}

In Fig.~\ref{fig:potential_evolution_nodiss}, we show the time evolution of the eigenstate occupation as given by the QSD equation without dissipation.
The lowest three levels become equally occupied regardless of the energy gaps
\footnote{
For the free case, the second and the fourth excited states are used because of the degeneracy of positive and negative momentum states.
}.
It is expected that not only these levels but all the levels are eventually occupied equally if the quantum dissipation is neglected.
We also observe that the effect of quantum dissipation at early time strongly depends on the external potential.
This dependence can be understood by analyzing the Lindblad equation as follows.
The initial decay rate of an eigenstate $\psi_i$ of the Hamiltonian is given by
\begin{align}
\Gamma_i &= -2\sum_{n}\left(
\langle L_n\rangle_{\psi_i}\langle L_n^{\dagger}\rangle_{\psi_i}
-\langle L_n^{\dagger}L_n\rangle_{\psi_i}
\right).
\end{align}
Using the Lindblad operators in Eq.~\eqref{eq:QCDLindblad}, we obtain
\begin{align}
\Gamma_i
&= D(0) - \int d^3xd^3y D(x-y)n_{\psi_i}(x) n_{\psi_i}(y) \\
& \quad +\frac{\nabla^2 D(0)}{4MT} + \frac{\nabla^4 D(0)}{64M^2T^2}
+\frac{\nabla_i\nabla_j D(0)}{16M^2T^2} \langle \nabla_i\nabla_j\rangle_{\psi_i}, \nonumber
\end{align}
in which the first (second) line arises from thermal fluctuations (dissipation).
With the Gaussian approximation \eqref{eq:dgaussian} for $D(x)$, we get for our one-dimensional model
\begin{align}
\label{eq:decayrate}
\frac{\Gamma_i}{\gamma} &= 1 - \int dxdy \exp\left[-\frac{(x-y)^2}{l_{\rm corr}^2}\right]n_{\psi_i}(x) n_{\psi_i}(y) \nonumber \\
& \quad -\frac{1}{2MTl_{\rm corr}^2} +\frac{3}{16M^2T^2l_{\rm corr}^4} 
- \frac{\langle\nabla^2\rangle_{\psi_i}}{8M^2T^2l_{\rm corr}^2}.
\end{align}
With the values from Table.~\ref{tbl:setup}, the effect of dissipation (the second line in Eq.~\eqref{eq:decayrate}) is $\simeq -0.048 - 0.125\langle\nabla^2\rangle_{\psi_i} /M^2$.
The ground state wave function yields
$\langle\nabla^2\rangle_{\psi_0}/M^2\simeq 0, -0.02, -0.08$ for the free
case, the harmonic potential, and the regularized Coulomb potential, respectively.
Therefore, the effect of dissipation in this case ranges from $-0.048$ to $-0.038$ and only slightly depends on the potential. 

On the other hand, the decay rate due to thermal fluctuation is quite sensitive to the size of the wave function as can be seen from the first line in Eq.~\eqref{eq:decayrate}.
In fact, the wave function size is $M\Delta x\simeq 3.5, 1.9 < 10=Ml_{\rm corr}$ for the ground states of the harmonic and the regularized Coulomb potentials respectively.
If the size is smaller than the correlation length $l_{\rm corr}$, the decay rate in the absence of dissipation is already comparatively small so that the relative importance of dissipation increases.

\section{Conclusion}
\label{sec:conclusion}
In this paper, we investigate how quantum dissipation influences the time evolution of the density matrix of a heavy quark in the quark-gluon plasma (QGP).
The master equation for heavy quark systems in the QGP has been obtained in the Lindblad form \cite{Akamatsu:2014qsa, lindblad1976generators} and thus possesses particularly useful properties: The density matrix $\rho$ stays hermitian, remains correctly normalized and positive during the time evolution.
We solve this Lindblad equation for a single heavy quark by stochastic unraveling.
Applying the approach of quantum state diffusion (QSD) \cite{gisin1992quantum} to our Lindblad equation, we derive a nonlinear stochastic Schr\"odinger equation for the heavy quark wave function.

Subsequently we solve the QSD equation for a heavy quark in simple settings, i.e. in one spatial dimension and with or without external potentials (harmonic and regularized Coulomb potentials).
We found that in both cases the density matrix relaxes to $\rho_{\rm eq}\propto e^{-H/T}$ within statistical errors.
This property is expected from a constraint in the Lindblad equation introduced by the fluctuation-dissipation theorem for the QGP sector but was not explicitly guaranteed by the Lindblad equation itself (see Appendix \ref{app:steadystate}).
We also found that the relaxation process strongly depends on the initial condition so that it is not captured by a simple rate equation.

As a further topic we study the effect of quantum dissipation by switching off the dissipative terms in the QSD equation.
Without the dissipative terms, the heavy quark is overheated because it only receives energy from the thermal medium which is not dissipated back.
It is shown that the importance of dissipation, as compared to the thermal fluctuations, strongly depends on the wave function size.
The relative importance of dissipation increases when the wave function is small.

In the future, we plan to extend our analysis to the description of heavy quarkonium in the QGP.
In that case, not only thermal fluctuations but also dissipation takes place nontrivially because the collisions of a heavy quark and those of a heavy antiquark interfere with each other.
It would be interesting and phenomenologically relevant to study the effects of quantum dissipation in that case.
The computations for quarkonium in three spatial dimensions and in evolving fluid background for heavy-ion collisions will be one of the ultimate goals of our project.

\begin{acknowledgments}
The work of Y. A. is partially supported by JSPS KAKENHI Grant Number JP18K13538. M. A. is supported in part by JSPS KAKENHI Grant Number JP18K03646.
Y.A. thanks the DFG Collaborative Research Centre SFB 1225 (ISOQUANT) for hospitality during his stay at Heidelberg University and A.R. was supported by SFB 1225 in full.
Y.A. also thanks T. Hirano for recommending \cite{percival1998quantum}.
\end{acknowledgments}

\ \\

\appendix
\section{Approximate steady state solution of the Lindblad equation}
\label{app:steadystate}
First we give a few algebraic relations of the Lindblad operators in Eq.~\eqref{eq:QCDLindblad}:
\begin{subequations}
\begin{align}
L_k \hat p &= (\hat p-k)L_k, \quad
L_k^{\dagger} \hat p = (\hat p+k)L_k^{\dagger},\\
L_k L_k^{\dagger} &= \frac{\tilde D(k)}{2V}\left(1-\frac{k\hat p}{4MT} + \frac{k^2}{8MT}\right)^2,\\
L_k^{\dagger} L_k &= \frac{\tilde D(k)}{2V}\left(1-\frac{k\hat p}{4MT} - \frac{k^2}{8MT}\right)^2.
\end{align}
\end{subequations}
The steady state solution $\rho_{\rm eq}(\hat p)$ of the Lindblad equation for a heavy quark without external potential $V_{\rm ext}=0$ satisfies
\begin{align}
0&=\sum_k\left(
2L_k\rho_{\rm eq}(\hat p) L_k^{\dagger} - L_k^{\dagger}L_k\rho_{\rm eq}(\hat p) - \rho_{\rm eq}(\hat p) L_k^{\dagger} L_k
\right)\nonumber \\
&=2\sum_k\left(
\rho_{\rm eq}(\hat p - k) L_kL_k^{\dagger} - \rho_{\rm eq}(\hat p) L_k^{\dagger} L_k
\right)\nonumber \\
&=2\sum_k\left(
\rho_{\rm eq}(\hat p - k) L_kL_k^{\dagger} - \rho_{\rm eq}(\hat p) L_{-k}^{\dagger} L_{-k}
\right).
\end{align}
In a collision with momentum transfer $k$, the on-shell energy of the heavy quark changes by an amount,
\begin{align}
\Delta E_{k} = \frac{p^2}{2M} - \frac{(p-k)^2}{2M} = \frac{2kp - k^2}{2M}.
\end{align}
In the momentum representation, $L_k L_k^{\dagger}$ and $L_{-k}^{\dagger} L_{-k}$ are given by
\begin{subequations}
\begin{align}
L_k L_k^{\dagger} &= \frac{\tilde D(k)}{2V}\left(1-\frac{\Delta E_k}{4T}\right)^2, \\
L_{-k}^{\dagger}L_{-k} &= \frac{\tilde D(k)}{2V}\left(1+\frac{\Delta E_k}{4T}\right)^2.
\end{align}
\end{subequations}
Then detailed balance in reactions between $p\leftrightarrow p-k$ dictates
\begin{align}
\frac{\rho_{\rm eq}(p)}{\rho_{\rm eq}(p-k)} &= \frac{L_k L_k^{\dagger}}{L_{-k}^{\dagger}L_{-k}}\nonumber \\
&\simeq 1-\frac{\Delta E_k}{T} + \frac{1}{2}\left(\frac{\Delta E_k}{T}\right)^2 + \cdots \nonumber \\
&\simeq e^{-\Delta E_k/T}.
\end{align}
Therefore up to the order $\mathcal O[(\Delta E_k/T)^3] \ll 1$, the steady state solution is the Boltzmann distribution
\begin{align}
\rho_{\rm eq}(p) \propto e^{-\frac{p^2}{2MT}}.
\end{align}
The above approximation is reliable for $\Delta E_k/T \ll 1$.
For $p\sim Mv$,  $\Delta E_k/T \sim v/Tl_{\rm corr} + 1/MTl_{\rm corr}^2\sim v+0.1$ so that our approximation is valid for $v \ll 1$.
Therefore, it is surprising that the numerical result in FIG. \ref{fig:free_pdist} is well approximated by the Boltzmann distribution for as large as $p\sim 1.5M$.

\bibliography{paper_v7}

\end{document}